\documentclass[aps,prb,reprint,superscriptaddress,showpacs]{revtex4-1} 

\usepackage{amsmath}     
\usepackage{amssymb}     
\usepackage{graphicx}    
\usepackage{hyperref}    

\begin{document}
\title{Surface states and local spin susceptibility in doped three-dimensional topological insulators with odd-parity superconducting pairing symmetry} 

\begin{abstract}
We investigate characteristic features in the spin response of doped three-dimensional topological insulators with odd-parity unequal-spin superconducting pairing, which are predicted to have gapless Majorana surface modes. These Majorana modes contribute to the local spin susceptibility, giving rise to a characteristic temperature behavior of the Knight shift and the spin-lattice relaxation time in magnetic resonance experiments. Because of their different decay lengths, the Majorana modes can be observed and clearly distinguished from the Dirac modes of the topological insulator by local probes, which allow for a depth-controlled study of the electron spins on the nanometer length scale. 
\end{abstract}

\author{Bj\"orn Zocher}
\affiliation{Institut f\"ur Theoretische Physik, Universit\"at Leipzig, D-04103 Leipzig, Germany}
\affiliation{Max Planck Institut f\"ur Mathematik in den Naturwissenschaften, D-04103 Leipzig, Germany}

\author{Bernd Rosenow}
\affiliation{Institut f\"ur Theoretische Physik, Universit\"at Leipzig, D-04103 Leipzig, Germany}

\date{\today}

\pacs{73.20.At, 74.20.Rp, 74.25.Ha, 74.25.nj}

\maketitle

\section{Introduction}
\label{sec:introduction}

Topological insulators (TIs) are time-reversal-invariant systems with gapped bulk and protected massless Dirac modes at the surface.\cite{HK2010,QZ2011} Very soon after the discovery of TIs, theorists generalized this concept to topological superconductors, which are characterized by a superconducting (SC) gap in the bulk and protected gapless Majorana surface modes.\cite{HK2010,QZ2011,SR2008,S2010,QH2009,LT2010}  Semiconductors like the bismuth chalcogenides with strong spin-orbit coupling and a Fermi surface centered at the time-reversal-invariant momentum, are of particular interest because of their single helical Dirac cone at the surface.\cite{ZL2009} Copper-doped Bi$_2$Se$_3$ is an unconventional superconductor~\cite{HW2010,WX2010,DS2011,KS2011} with non-trivial surface states and a band structure similar to that of Bi$_2$Se$_3$ but with shifted chemical potential, reduced Fermi velocity, and enlarged surface dispersion.\cite{WX2010}

By now, the surface states in Cu$_x$Bi$_2$Se$_3$ have been probed by photoemission\cite{WX2010} and point contact spectroscopy.\cite{SK2011,TL2011,Chen+12,Levy+12,Peng+13} While the experiments \cite{SK2011,TL2011,Chen+12} find evidence for midgap states and hence topological superconductivity, no such states were found in Refs. \onlinecite{Levy+12,Peng+13}. As a result, the question about the pairing symmetry of  Cu$_x$Bi$_2$Se$_3$ cannot be clearly answered from point contact spectroscopy at the moment, and results obtained by complementary experimental techniques are desirable. Nuclear magnetic resonance (NMR) and quadrupole resonance, as well as the electron and muon spin resonance ($\mu$SR) are another class of powerful techniques to investigate the electronic properties locally. The Knight shift for example is determined by the static spin susceptibility $K \sim \chi_s(\textbf{q}=0,\omega=0)$, which is directly connected to the spin structure of the SC pairing. In conventional $s$-wave superconductors with spin-singlet pairing, the Knight shift is significantly reduced and vanishes for $T=0$ because spins pair up and longitudinal spin excitations cost the pair-breaking energy $2\Delta$. However, in superconductors with strong spin-orbit coupling the spin susceptibility is suppressed as compared to the normal state but does not vanish for $T=0$ due to coupling between up and down spins.\cite{A1959} In this paper, we study characteristic features in the spin response of odd-parity pairing in doped TIs and predict clear signatures for the above resonance techniques. 

Fu and Berg\cite{FB2010} showed that strong spin-orbit-coupled bands indeed favor an odd-parity interorbital unequal-spin pairing.\cite{HL2011,YY2012} To gain insight into its topological non-trivial nature, we map this pairing Hamiltonian onto the conduction band, which yields an effective time-reversal invariant $p \pm ip$ pairing in three dimensions. Because of this topology, there is a pair of Majorana zero-energy modes (MZM) located at each surface and protected by time-reversal symmetry. Additionally there are unconventional surface Andreev bound states (SABSs) originating from the band inversion as shown in Ref. \onlinecite{HF2011} for a linear $k \cdot p$ model. In addition to the linear momenta, we here consider quadratic momentum terms, which determine the energy range of coexistence between Dirac modes and unconventional SABSs, and which may give rise to another species of zero-energy SABS. The main motivation for introducing the quadratic terms is the possibility to investigate the competition between the different surface states and the bulk. The coexistence of the MZMs and the SABSs originating from the band inversion\cite{HF2011} gives rise to two characteristic length scales. The Dirac modes decay on the nm scale $\xi_0$\cite{LY2009,LZ2010,LS2010} whereas the decay length $\xi_1$ for the MZMs is hundreds of nm.\cite{HF2011} Hence, the local spin susceptibility shows different characteristic behavior in the bulk, at the surface, and within $\xi_1$ into the bulk. 

The origin for the spin response of the SABS is the helical spin structure of the quasiparticle dispersion. In superconductors with unequal-spin-pairing symmetry, the creation operator for a Bogoliubov quasiparticle contains a particle component with, say, momentum $\textbf{k}$ and spin $\uparrow$, and a hole component with momentum $-\textbf{k}$ and spin $\downarrow$. Since a hole with spin $\downarrow$ is the absence of a particle with spin $\downarrow$, this removal of a spin $\downarrow$ coherently adds up with the addition of a spin $\uparrow$ particle, and gives rise to a spinful quasiparticle. Since, the direction of the quasiparticle spin depends on the direction of momentum $\textbf{k}$, the states at $\textbf{k}=0$ are special, because the direction of their momentum, and as a consequence the direction of their spin, is undetermined.  Because of time-reversal symmetry, on each surface two modes reside at zero energy and span a two-dimensional vector space. One special basis of this vector space is determined by quasiparticle operators, which satisfy the Majorana criterion, $\gamma_\pm=\gamma_\pm^\dagger$. These basis vectors are invariant under time-reversal symmetry\cite{HF2011} and thus the Majorana operators itself are "spinless". Similarly, we choose another basis for the quasiparticle operators as superpositions of the Majorana operators, $\gamma_\sigma=\gamma_+ +\sigma i\gamma_-$. In contrast to the Majorana basis discussed above, these operators are fermion operators with $\gamma_\uparrow ^\dagger =\gamma_\downarrow$ and thus non-Hermitian. Furthermore, these operators transform into each other under time-reversal symmetry and in particular they are fully spin polarized with spin $\sigma$ in the $z$ direction. We conclude that without loss of generality the zero-energy SABS can be understood as a spinful quasiparticle. From this perspective, it becomes clear, that the MZMs contribute to the magnetic properties of the topological superconductor. In particular, from the spin polarization one can easily see that for a Zeeman field in $z$ direction, the spinful basis is an eigenbasis of the Zeeman coupling and the MZMs acquire a finite energy proportional to the magnetic field. In contrast, in a spin-polarized $p$-wave topological superconductor, the combination of a spin $\uparrow$ particle component and a spin $\uparrow$ hole component indeed has zero net spin and would not contribute to the spin response.\cite{QH2009}

Conventional bulk NMR can distinguish between competing pairing symmetries by the characteristic temperature dependence of the Knight shift and the spin-lattice relaxation rate. We propose that NMR in thin films of thickness $L \sim 500$ nm or depth controlled probes\cite{MF2004,SB2000} allow to clearly determine the pairing symmetry and investigate the MZMs. Our work is motivated by Cu$_x$Bi$_2$Se$_3$, however, our results are more generally relevant for other inversion symmetric materials such as the ternary chalcogenides\cite{LM2010} and the PbTe class.\cite{HL2012,SR2012} Furthermore, our findings for doped TIs are complementary to the superfluid $^3$He-$B$ phase\cite{CS2009,NH2009} where spin relaxation reflects the gapless Majorana nature. 

Our paper is organized as follows. In Sec. \ref{sec:model_system}, we introduce the model system for the TI and competing superconducting pairing symmetries. We continue in Sec. \ref{sec:magnetic_response} with the study of the spin response where we concentrate on the real part of the longitudinal spin susceptibility, which yields the Knight shift, and the imaginary part of the transverse spin susceptibility, which determines the spin lattice relaxation rate. In Sec. \ref{sec:experimental_detection_schemes}, we compare the spin response for the various pairing symmetries and predict magnetic resonance experiments to observe the unconventional SABS in topological superconductors. We summarize our results in Sec. \ref{sec:conclusion}.

\section{Model system}
\label{sec:model_system}

\subsection{Hamiltonian for the superconducting TI}

In the following, we consider doped three-dimensional TIs described by the low-energy $k \cdot p$ Hamiltonian $H_{TI} =\sum_{\textbf{k}} C_{\textbf{k}}^\dagger  \mathcal{H}_{TI}(\textbf{k})C_{\textbf{k}}$,\cite{ZL2009,FB2010} where
%
\begin{equation}
\mathcal{H}_{TI}(\textbf{k})= m_0(\textbf{k})\sigma_x +v_z k_z\sigma_y +v (k_x s_y-k_y s_x) \sigma_z -\mu
\label{eqn:HTI}
\end{equation}
%
with $m_0(\textbf{k}) =m+ B_1 k_z^2 +B_2 (k_x^2+k_y^2)$, $C_{\textbf{k}}=(c_{i\textbf{k}s})_{i,s}$, $B_1,v_z>0$, and $m<0$. Here, the $s_i$ ($\sigma_i)$ denote the Pauli matrices for the spin (orbital) degree of freedom, and the operators $c_{i\textbf{k}s}$ annihilate an electron in orbital $i$, with momentum $\textbf{k}$, and spin $s$. The unitary operator $\mathrm{exp}(i\phi s_z/2)$, where $\phi$ is the azimuthal angle $(k_x,k_y)$, transforms Hamiltonian Eq.~\eqref{eqn:HTI} onto $\mathcal{H}_{TI}(k_x,0,k_z)$, which reflects the rotational symmetry around the $z$ axis. The doped charge density determines the chemical potential $\mu$, and the Fermi surface is given by the equation $\mu^2=m_0^2(\textbf{k}_F)+v_z^2k_{F,z}^2+v^2(k_{F,x}^2+k_{F,y}^2)$. 

The conduction band of the doped TI is described by the operators $\alpha_{\textbf{k},\tau}=\sum_{s,\sigma} \psi_{\tau}^{s,\sigma}(\textbf{k})c_{\sigma,\textbf{k},s}$ with 
%
\begin{equation}
\psi_{\tau}(\textbf{k})=\frac{1}{2\sqrt{E}}
\begin{pmatrix}
\sqrt{E+vk\tau}e^{-i\chi/2}\\
\sqrt{E-vk\tau}e^{i\chi/2}
\end{pmatrix}_\sigma \otimes
\begin{pmatrix}
e^{-i(\phi/2 +\tau\pi/4)}\\
e^{i(\phi/2 +\tau\pi/4)}
\end{pmatrix}_s,
\end{equation}
%
$e^{i\chi}=[m_0(\textbf{k})+iv_zk_z]/\sqrt{m_0^2(\textbf{k})+v_z^2k_z^2}$, and $E=\sqrt{m_0^2(\textbf{k})+v_zk_z^2+v^2k^2}$. The operators $\alpha_{\textbf{k},\tau}$ satisfy the condition $\alpha_{(k,\phi+2\pi,k_z),\tau}=-\alpha_{(k,\phi,k_z),\tau}$, i.e., the operators are $4\pi$ periodic under rotation in momentum space. Because of the helical band structure, the operators $\alpha_{\textbf{k},\tau}$ and $\alpha_{-\textbf{k},\tau}$ transform into each other under time reversal while $\mathcal{P}\alpha_{\textbf{k},\tau}\mathcal{P}^{-1}=\tau\alpha_{-\textbf{k},-\tau}$. Here, the parity operator $\mathcal{P}$ transforms $c_{1\textbf{k}s}$ into $c_{2-\textbf{k}s}$ and vice versa.

%
\begin{figure*}[t]
\centering
  \includegraphics[width=1.3\columnwidth]{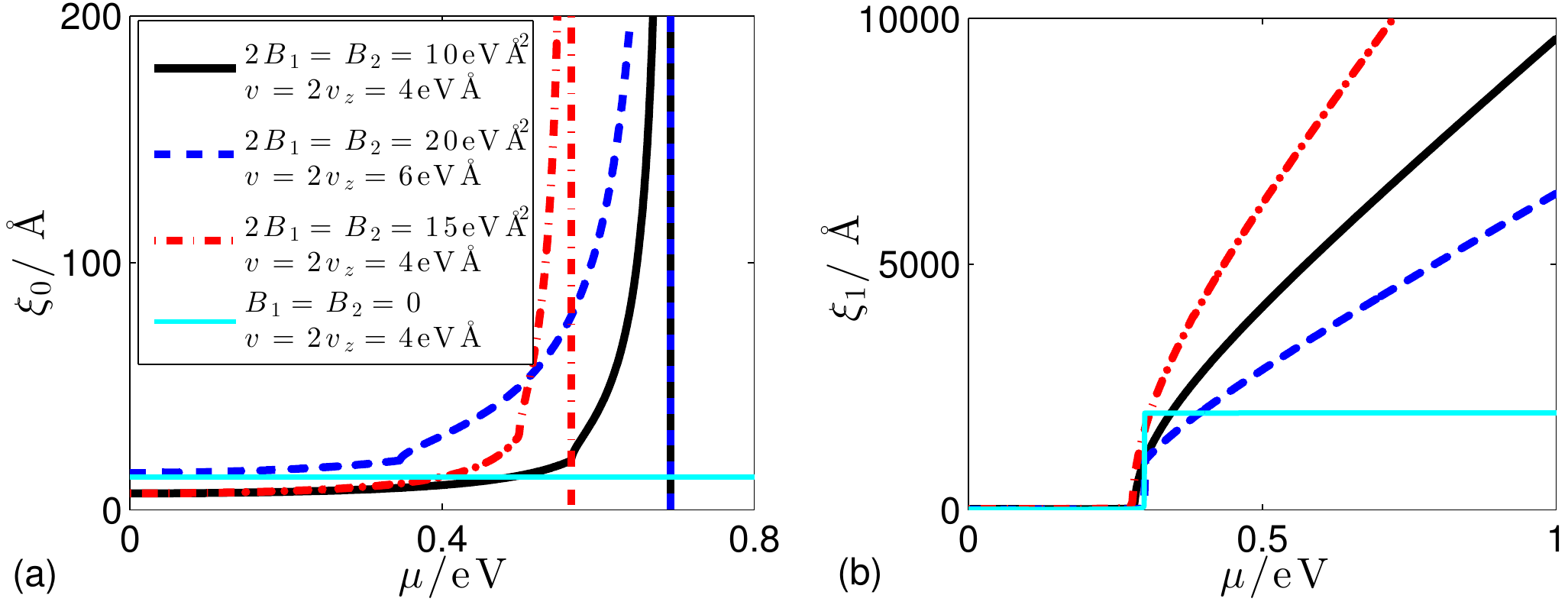}
\caption{(Color online) Decay lengths of (a) the Dirac mode of the TI and (b) the zero-momentum MZM of the odd-parity interorbital superconductor obtained from the analytical expressions for the wave functions for the semi-infinite TI with $\Delta=1$ meV and $m=-0.3$ eV. The lines are defined in panel (a). }
\label{fig:decay}
\end{figure*}
%

Possible pairing terms depend on the specific mechanism and the lattice symmetry. For pairing induced by on-site density-density interactions, Fu and Berg\cite{FB2010} showed that the spin-orbit coupled bands favor odd-parity interorbital unequal-spin pairing,  
%
\begin{equation}
H_{SC}=\Delta\sum_{\textbf{k}}\Big( c_{1\textbf{k}\uparrow}c_{2-\textbf{k}\downarrow}+c_{1\textbf{k}\downarrow}c_{2-\textbf{k}\uparrow}\Big)+H.c. ,
\label{eqn:HSC}
\end{equation}
%
where odd-parity pairing denotes $\mathcal{P}H_{SC}\mathcal{P}^{-1}=-H_{SC}$ under parity transformation. If the pairing is induced by long-range interactions, such as the electron-phonon interaction, other unequal-spin pairing channels are also possible.\cite{HL2011} Additionally, we consider
%
(i)    even-parity, intraorbital pairing
\begin{equation}
H_1=\Delta_1 \sum_{\textbf{k}}\Big( c_{1\textbf{k}\uparrow}c_{1-\textbf{k}\downarrow}+c_{2\textbf{k}\downarrow}c_{2-\textbf{k}\uparrow}\Big)+H.c.,
\label{eqn:H01}
\end{equation}
(ii)   odd-parity, intraorbital pairing
\begin{equation}
H_2=\Delta_2 \sum_{\textbf{k}}\Big( c_{1\textbf{k}\uparrow}c_{1-\textbf{k}\downarrow}-c_{2\textbf{k}\downarrow}c_{2-\textbf{k}\uparrow}\Big)+H.c.,
\label{eqn:H02}
\end{equation} 
and (iii)  even-parity, interorbital pairing
\begin{equation}
H_3=\Delta_3\sum_{\textbf{k}}\Big( c_{1\textbf{k}\uparrow}c_{2-\textbf{k}\downarrow}-c_{1\textbf{k}\downarrow}c_{2-\textbf{k}\uparrow}\Big)+H.c.
\label{eqn:H03}
\end{equation}
%

\subsection{Doped TI with odd-parity interorbital pairing}

In this section, we investigate the doped TI with odd-parity interorbital pairing Eq.~\eqref{eqn:HSC}. To study the effect of spin-orbit coupling on the pairing symmetry, we project Eq.~\eqref{eqn:HSC} onto the basis $(\alpha_{\textbf{k},\tau})$ spanned by the conduction band of the bulk TI Eq.~\eqref{eqn:HTI}. For $\mu > -m \gg \Delta$, this yields the effective pairing Hamiltonian 
%
\begin{equation}
H_{SC} \approx  i\Delta \sum_{\textbf{k}} \frac{v_zk_z +ivk \frac{m_0(\textbf{k})}{\mu}}{E_0(\textbf{k})}  \alpha_{\textbf{k},+}\alpha_{-\textbf{k},-} +H.c. ,
\label{eqn:triplet}
\end{equation}
%
where $E_0^2(\textbf{k})=v_z^2k_z^2+v^2k^2$ and $k^2=k_x^2+k_y^2$. The effective pairing Hamiltonian is exact in first order in $\Delta/\mu$ and yields a gapped bulk excitation spectrum with quasiparticle gap $2\Delta E_0(\textbf{k})/\mu$. The corresponding Bogoliubov-de Gennes equations are in the same universality class as the ones for two copies of spinless superconductors with opposite chirality, which is known to be a time-reversal-invariant topological superconductor with MZMs if the chemical potential lies within the conduction band.\cite{RG2000,QH2009} From this analogy, we expect to find a Kramers pair of MZMs for $k=0$ and additionally a pair of zero-energy SABSs whenever $m_0(\textbf{k}_F)=0$. In contrast, the bulk single-particle excitation spectrum is always fully gapped with weakly momentum dependent gap $\Delta\sqrt{1-m_0^2(\textbf{k}_F)/\mu^2}>0$. 

The Hamiltonian Eq.~\eqref{eqn:HTI} confined in $z$ direction has a two-dimensional helical massless Dirac cone at the surface.\cite{ZL2009} However, we also know that the effective pairing Hamiltonian Eq.~\eqref{eqn:triplet} yields a pair of helical MZMs at the time-reversal-invariant momentum $k=0$.\cite{QH2009} Hence, we obtain two species of surface states originating from the band inversion and the $p \pm ip$ pairing, respectively. We find these states by replacing $k_z \rightarrow -i \partial_z$ and solving the corresponding Schr\"odinger equation with boundary condition $\sigma_z \psi(z=0)=\psi(z=0)$, which describes the vanishing of the wave function for orbital 2 at the surface. This boundary condition is justified by the layered atomic structure of the TI where orbital 2 is underneath the orbital 1 surface layer. 

For the normal conducting Hamiltonian Eq.~\eqref{eqn:HTI}, we obtain the dispersion $E_{D,\tau}(k)=vk$ and the surface wave functions 
%
\begin{equation}
\psi _{D, \tau}(z,k,\phi)=\sqrt{\nu/v_z} \,      ( e^{ \nu z /v_z}, 0)_{\sigma}     \otimes     (1, i\tau e^{i\phi})_s,
\label{eqn:EVTI_edge}
\end{equation}
%
where $\nu=\big(1-\sqrt{1+4B_1(m+B_2k^2)/v_z^2}\big)/2B_1$. From this expression, we find the decay length $\xi_0=2B_1/v_z$ for $k^2< (-m-v_z^2/4B_1)/B_2 $ and $\xi_0=2B_1/v_z/\big[1-\sqrt{1+4B_1(m+B_2k^2)/v_z^2}\big]$ for $(-m-v_z^2/4B_1)/B_2<k^2< -m/B_2 $. For $k^2\rightarrow -m/B_2$, $\xi_0$ diverges and the surface states become bulk states as shown in Fig.~\ref{fig:decay}(a). 

%
\begin{figure*}[t]
\centering
  \includegraphics[width=1.3\columnwidth]{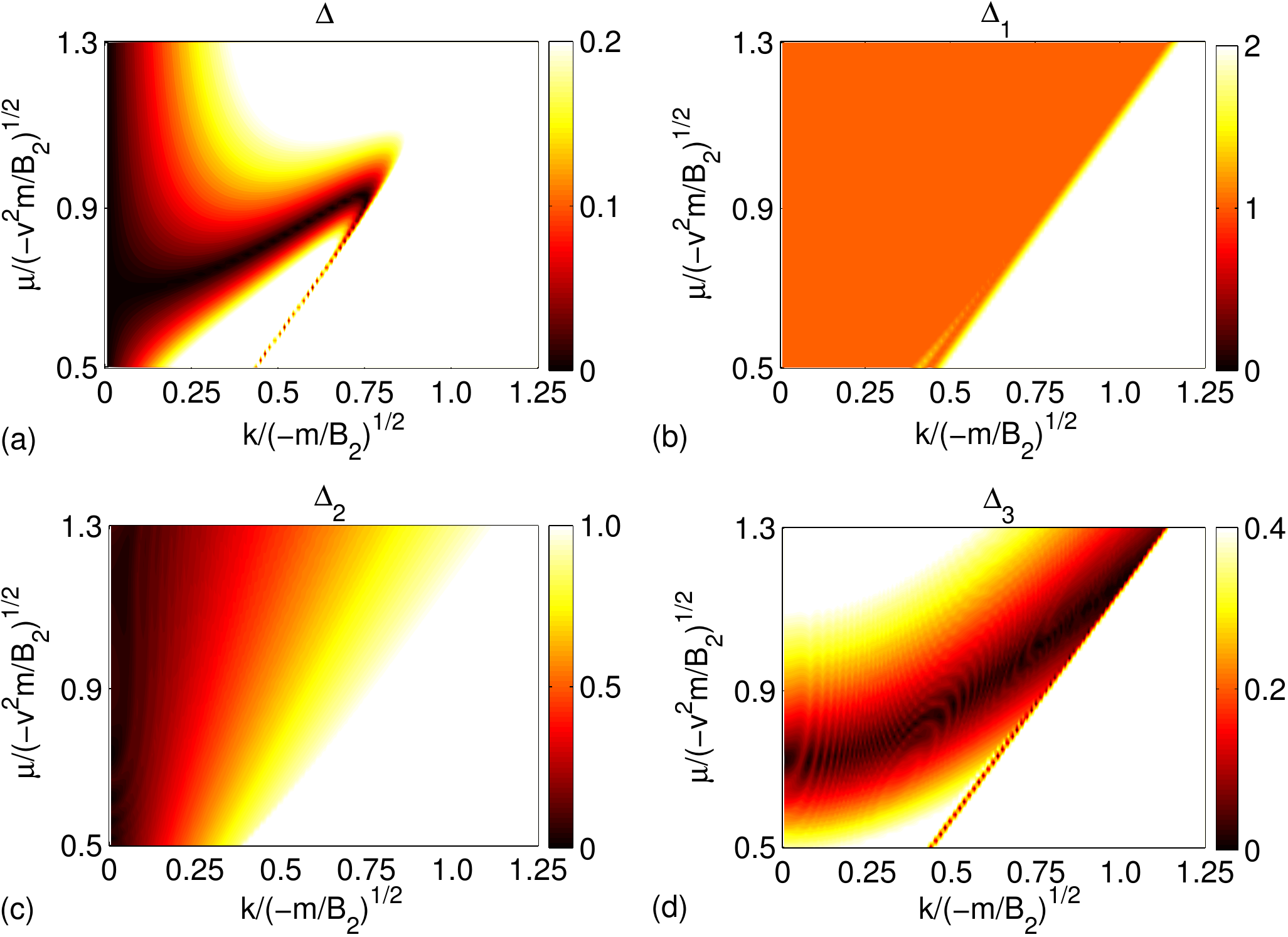}
\caption{(Color online) Lowest quasiparticle energy in units of $\Delta_i$ for a thin film of thickness $L=240$ nm confined in the $z$ direction and with (a) $\Delta$, (b) $\Delta_1$, (c) $\Delta_2$, and (d) $\Delta_3$ pairing as function of in-plane momentum $k$ and chemical potential $\mu$. Note the different color scales.}
\label{fig:le_spectrum}
\end{figure*}
%
Similarly as for the Dirac modes, we find the zero-momentum MZMs by solving the Schr\"odinger equation. Thus, we obtain the Majorana dispersion
%
\begin{equation}
E_M(k)\approx vk \frac{\Delta(m+B_1k_{F,z}^2)}{\mu^2}
\end{equation} 
%
near $k=0$. The MZMs decay on a characteristic length $\xi_1$ as shown in Fig.~\ref{fig:decay}(b) where we plot $\xi_1$ as function of the chemical potential. For $\mu > |m|$, we find  $\xi_1(\mu)=v_{F,z}/\Delta$ with the Fermi velocity in the $z$ direction $v_{F,z}$. Typical values for $\xi_1$ are by a factor of $E_0/\Delta \sim 10^2$ larger than $\xi_0$. In contrast to the Dirac modes, the zero-momentum MZMs exist for all values of the chemical potential and enter the bulk for $\mu \rightarrow \infty$ only. For both species of surface states, the quadratic terms $B_1$ and $B_2$ significantly determine the behavior of the decay lengths. For $B_1=B_2=0$, we obtain $\xi_0=v_z/|m|$ and $\xi_1=v_z/\Delta$ whereas the decay lengths for $B_1,B_2>0$ strongly depend on the Fermi velocity and thus on $\mu$. We find that the MZMs are immune against band bending effects due to near-surface electrostatic potential variations\cite{B2012} since the characteristic length for band bending effects is nm and thus much smaller than the decay length $\xi_1$. Furthermore, the MZMs are robust against moderate non-magnetic impurity scattering since pair-breaking effects are suppressed by an approximate chiral symmetry in the spin-orbital locked band structure.\cite{KF2012}

Depending on the doped charge density, the SC state could occur with the chemical potential either in the region where the Dirac modes are separated from the bulk conduction band\cite{WX2010} or where only the bulk states remain. Since the numerator of the effective pairing Hamiltonian Eq.~\eqref{eqn:triplet} vanishes for the Dirac modes with $k_z \rightarrow -i\nu/v_z$ and $\mu=vk$, they are not gapped by $H_{SC}$ and yield a ring structure of zero-energy SABSs. Higher order terms which couple valence and conduction band, do not change the character of the surface modes qualitatively. The authors of Ref.~\onlinecite{HF2011} showed that a branch of SABSs connects the MZM and the Dirac modes due to the mirror helicity of the Hamiltonian, which here shows up as the sign of the mass $m_0(\textbf{k}_F)$ in Eq. \eqref{eqn:triplet}. This mass is negative near the bottom of the conduction band and changes sign for $\mu^2=-mv_z^2/B_1$. However, in contrast to the zero-momentum mode, the finite-momentum modes are not MZMs in the sense that they are equal superpositions of electron and hole creation operators such that $\gamma=\gamma^\dagger$. Instead, the Bogoliubov operators for these finite-momentum modes satisfy $\gamma_{\textbf{k}}^{}=\gamma_{-\textbf{k}}^\dagger$ with a finite electron-hole imbalance.

In Fig.~\ref{fig:le_spectrum}(a) we plot the energy of the surface states as function of momentum and chemical potential where we see a Kramers pair of zero-momentum MZMs for all $\mu$ and depending on the chemical potential, we find three regimes, which can be distinguished by the number of additional finite-momentum zero-energy modes. For small chemical potentials, $\mu^2<-mv_z^2/B_1,-mv^2/B_2$, the mass term is negative, and hence, both the MZM and the branch of zero-energy SABS originating from the Dirac mode exist. On the other hand, for large chemical potentials, $-mv_z^2/B_1, -mv^2/B_2 <\mu^2$, the mass term $m_0(\textbf{k}_F)$ is positive and only the zero-momentum MZM exists. In the regime of intermediate $\mu$, we distinguish two cases $-mv_z^2/B_1\gtrless-mv^2/B_2$. For $-mv_z^2/B_1 \le \mu^2< -mv^2/B_2$, there is a momentum $\textbf{k}_F$ such that the mass term $m_0(\textbf{k}_F)$ vanishes. As shown in Fig.~\ref{fig:le_spectrum}(a) this yields another species of zero-energy SABS emerging at $k=0$, which now carries the negative velocity from the band inversion as expected from Eq.~\eqref{eqn:triplet} and moves towards the Dirac mode with increasing $\mu$ and is located at the in-plane momentum
%
\begin{equation}
k \approx \sqrt{\frac{B_1\mu^2-|m|v_z^2}{B_1v^2-B_2v_z^2}}. 
\label{eqn:moving}
\end{equation}
%
For $\mu=vk$, both finite-momentum SABS meet and gap out for $\mu^2 \rightarrow -mv^2/B_2$. In contrast, for $-mv^2/B_2 <\mu^2< -mv_z^2/B_1$, the Dirac modes disappeared in the bulk and is replaced by the unconventional SABS discussed by Hsieh and Fu\cite{HF2011} while $m+B_1k_{F,z}^2$ is negative. Moreover, with increasing chemical potential, this finite-momentum zero-energy SABS moves towards $k=0$ and disappears for $\mu^2 \rightarrow -mv_z^2/B_1$. Thus, we conclude that the number of species of zero-energy SABS is even for $m^2<\mu^2<-mv_z^2/B_1$ and odd for $\mu^2>-mv_z^2/B_1$. Hsieh and Fu\cite{HF2011} do not find the competition between these different sectors since the parameters $B_1$ and $B_2$ in their lattice model are small and thus they only consider the small chemical potential regime with the zero-momentum MZM and one species of finite-momentum modes.

\subsection{Competing pairing symmetries}

In analogy to the study of the odd-parity interorbital pairing, we here investigate the effect of spin-orbit coupling on the competing pairing symmetries. Hence, we project Eqs.~\eqref{eqn:H01}--\eqref{eqn:H03} onto the basis $(\alpha_{\textbf{k},\tau})$ spanned by the conduction band of the bulk TI Eq.~\eqref{eqn:HTI}, which yields
%
\begin{subequations}
\begin{align}
H_1 &\approx \Delta_1 \sum_{\textbf{k},\tau} \alpha_{\textbf{k},\tau}\alpha_{-\textbf{k},\tau}+ H.c.,
\label{eqn:H1}\\
H_2 &\approx \Delta_2 \sum_{\textbf{k},\tau}\frac{\tau vk}{\mu}\alpha_{\textbf{k},\tau}\alpha_{-\textbf{k},\tau} +H.c.,
\label{eqn:H2}\\
H_3 &\approx \Delta_3 \sum_{\textbf{k},\tau}\frac{m_0(\textbf{k})}{\mu} \alpha_{\textbf{k},\tau}\alpha_{-\textbf{k},\tau}+H.c..
\label{eqn:H3}
\end{align}
\end{subequations}
These equal pseudospin pairing terms do not vanish because of the $4\pi$ periodicity of the operators $\alpha_{\textbf{k},\tau}$ under rotation in the $k_x$-$k_y$ plane. Even if these effective pairing symmetries are very similar, we observe differences in the effect on the Dirac modes and characteristic bulk behavior when the mass term $m_0(\textbf{k})$ changes sign. In Fig.~\ref{fig:le_spectrum}, we plot the energy of the lowest quasiparticle level for the superconductor confined in the $z$ direction as function of in-plane momentum $k$ and chemical potential $\mu$.  

For $\Delta_1$, the effective pairing Eq.~\eqref{eqn:H1} is intraorbital with the same amplitude for both pseudospins. Hence, we find conventional $s$-wave behavior for a metal with spin-orbit coupling. As shown in Fig.~\ref{fig:le_spectrum}(b), the single-particle excitation spectra in the bulk and at the surface are both fully gapped with gap $2\Delta_1$. The gap at the surface originates from the fact that the helical Dirac modes have only contributions from one orbital and the pairing is intraorbital. We find that even if the Dirac modes are gapped, they do not hybridize with the bulk states because there is no particle-hole mixing between the bulk and the surface and the gap at the surface arises from the pairing between the helical Dirac modes only. 

For $\Delta_2$-pairing [Fig.~\ref{fig:le_spectrum}(c)], the orbitals pair into singlets with a relative minus sign. Because of this relative minus sign, the effective pairing term Eq.~\eqref{eqn:H2} is linear in $k$ and vanishes for $k_x=k_y=0$, which gives rise to point nodes in the bulk single-particle excitation spectrum and a linear dispersion for $\epsilon \ll \Delta_2$. However, for the Dirac modes with $\mu=vk$, we find as for the even parity case a gap $2\Delta_2$. 

The effective pairing Hamiltonian for $\Delta_3$ is shown in Eq.~\eqref{eqn:H3} with a fully gapped bulk single-particle excitation spectrum for $m_0(\textbf{k}_F)\neq 0$ and a momentum dependent gap $2\Delta_3|m_0(k)|/\mu$. Since $m_0(k)$ changes sign when $k$ increases because of the band inversion at $k=0$, we find a gap closing and reopening as function of $\mu$, which is intricately related to the transition from the TI phase into the band insulator phase. Furthermore, $\Delta_3$ does not gap the Dirac modes for which $m_0(k,-i\nu/v_z)=\nu$ and the contributions from $\nu$ and $-\nu$ cancel. However, from the quasiparticle energies we see that the bulk spectrum shows nodes whenever $m_0(k)$ changes sign as shown in Fig.~\ref{fig:le_spectrum}(d). Similarly to the $\Delta$ case, we here find three qualitatively distinct sectors with transitions at $\mu^2=-mv_z^2/B_1$ and $\mu^2=-mv^2/B_2$. The origin of the nodes and distinct sectors is again the band inversion near $k=0$, which changes into the trivial band order along the gapless line in Fig.~\ref{fig:le_spectrum}(d). However, in contrast to the odd-parity interorbital case, here, the gapless modes are bulk modes. In our numerics, we only consider the case $\mu^2<-mv_z^2/B_1<-mv^2/B_2$, where the bulk is fully gapped and the Dirac modes are ungapped.

\section{Magnetic response}
\label{sec:magnetic_response}

In the following, we investigate signatures of the surface states in the local quasiparticle density of states (LDOS) and the local spin response of the doped TI confined in the $z$ direction. For the numerics we discretize the $z$ direction of Hamiltonian Eq.~\eqref{eqn:HTI} with surfaces at $z=0$ and $z=L$ by replacing $k_z\psi \rightarrow -i(\psi_{n+1}-\psi_{n-1})/2a$ and $k_z^2 \psi \rightarrow -(\psi_{n+1}+\psi_{n-1}-2\psi_n)/a^2$, where $z=na$. In our numerics, we consider a film of thickness of $L=240$ nm and with lattice constant $a=6$ \AA. Motivated by the Bi chalcogenides,\cite{QZ2011} we use the parameters $m=-0.3$ eV, $v=2v_z=4$ eV{\AA}, $B_2=2B_1=10$ eV {\AA}$^2$, $\mu=0.5$ eV, and $\Delta=3$ meV. Here, $B_2$ is reduced as compared to Bi$_2$Se$_3$ to guarantee the existence of the Dirac modes and its separation from the conduction band as found for Cu$_x$Bi$_2$Se$_3$.\cite{QZ2011,WX2010} From the above analysis we know that the zero-momentum MZMs and one finite momentum zero-energy SABS coexist in this regime. For abbreviation, we neglect the quadratic momentum terms in our analytical results where they mainly renormalize the Fermi velocity. 

The dynamical spin susceptibility is defined by
%
\begin{equation}
\chi_{ij}(\textbf{q},i\omega_n) =  \frac{1}{\mathcal{V}}\int^\beta_0  \left\langle   T_\tau  S^i_\textbf{q}(\tau) S^j_{-\textbf{q}}(0)   \right\rangle e^{i\omega_n\tau} d\tau , 
\label{eqn:chi1}
\end{equation}
where $S^{j}_{\textbf{q}}$ is the spin operator,
\begin{equation}
S^j_\textbf{q} = \frac{1}{\mathcal{V}}  \sum_{a,b=1,2}  \sum_\textbf{k} \sum_{s, s'}  c_{a\textbf{k}+\textbf{q} s}^\dagger  \frac{\sigma^j_{s s'}}{2}   c_{b \textbf{k} s'}^{}. 
\label{eqn:s1}
\end{equation}
%
In our numerics, we use the analytical continuation $i \omega_n \rightarrow \omega +i\delta$ with the broadening $\delta=\Delta/10$. 

%
\begin{figure*}[t]
\centering
  \includegraphics[width=1.3\columnwidth]{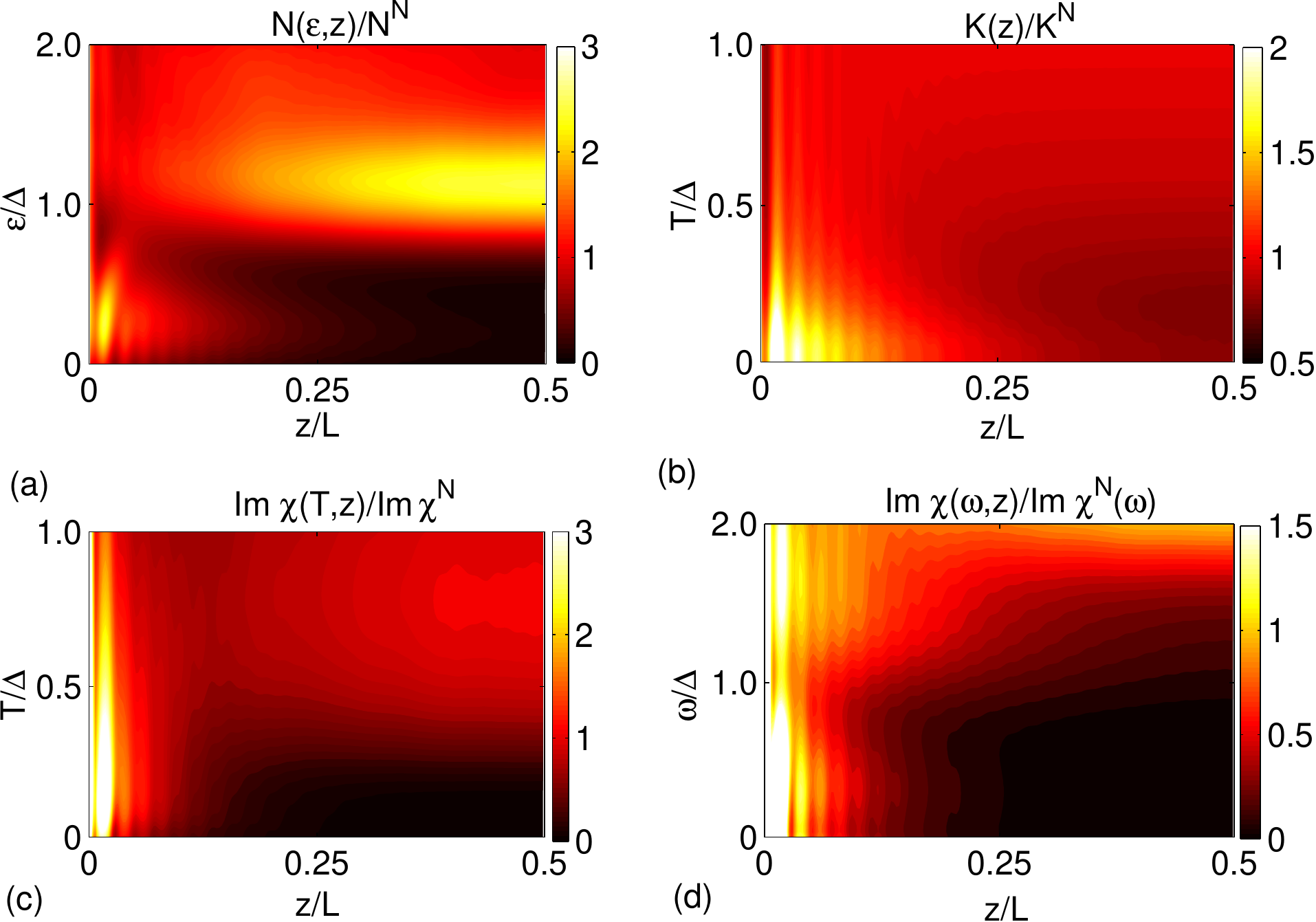}
\caption{(Color online) Spin response for a TI film of thickness $L=240$ nm with odd-parity interorbital $\Delta$ pairing. (a) Local density of states, (b) Knight shift, and imaginary part of the transverse spin susceptibility as function of (c) temperature and (d) excitation energy with $T=0$. All quantities are normalized to the normal-conducting bulk at $T=0$. }
\label{fig:TSC_4}
\end{figure*}
%

\subsection{Odd parity, interorbital pairing $\Delta$}

\subsubsection{Local density of states}

In Fig.~\ref{fig:TSC_4}(a) we plot the LDOS which shows two qualitatively distinct regions. In the bulk we obtain 
%
\begin{equation}
\mathcal{N}_B(\epsilon)=\frac{2\mu E_0(k_F) \epsilon}{\pi v^2v_z\sqrt{\epsilon^2-\frac{\Delta^2E_0^2(k_F)}{\mu^2}}}  \Theta \left(\epsilon-\frac{\Delta E_0(k_F)}{\mu} \right)
\end{equation}
%
with quasiparticle gap $2\Delta E_0(k_F)/\mu$ and sharp coherence peak, while there is a finite midgap LDOS at the surface. Depending on whether the Dirac modes already crossed the bulk band, we distinguish between $\mu^2>v^2|m|/B_2$ where only the MZMs appear with surface LDOS
%
\begin{equation}
\mathcal{N}_S(\epsilon,z \ge 0) \approx \epsilon \frac{\mu ^4 e^{-2z/\xi_1}}{\pi \xi_1 v^2 m^2 \Delta^2}\sin^2 \Big(\frac{zE_0}{v_z}\Big)
\end{equation}
%
and $\mu^2<v^2|m|/B_2$ where both the MZM and Dirac mode exist with
%
\begin{equation}
\mathcal{N}_S(\epsilon,z \ge 0) \approx \frac{\mu e^{-2z/\xi_0}}{\pi \xi_0 v^2}+  \frac{\epsilon\mu ^4 e^{-2z/\xi_1}}{\pi \xi_1 v^2 m^2 \Delta^2}\sin^2 \Big(\frac{zE_0}{v_z}\Big)
\label{eqn:Nloc}
\end{equation}
%
for $\epsilon \ll \Delta$ as shown in Fig.~\ref{fig:TSC_4}(a). The different surface states can be clearly distinguished by their decay lengths $\xi_0 \ll \xi_1$ and their energy dependencies. While the LDOS of the Dirac modes is almost constant as function of energy, the LDOS of the Majorana SABS strongly depends on energy with a linear increase for $\epsilon \ll \Delta$ and a peak for $\epsilon \lesssim \Delta/2$.\cite{HF2011} The origin for this very different energy behavior relies on the different energy scales. While the Dirac modes disperse on the scale of the band mass $m$, the Majorana SABSs disperse on the scale $\Delta \ll |m|$. Hence, on the energy scale $\Delta$ the Dirac modes show a constant LDOS as function of energy. Furthermore, the MZMs oscillate with a period $\lambda=v_z/E_0$, which is on the nm scale.

%
\begin{figure*}[t]
\centering
  \includegraphics[width=1.3\columnwidth]{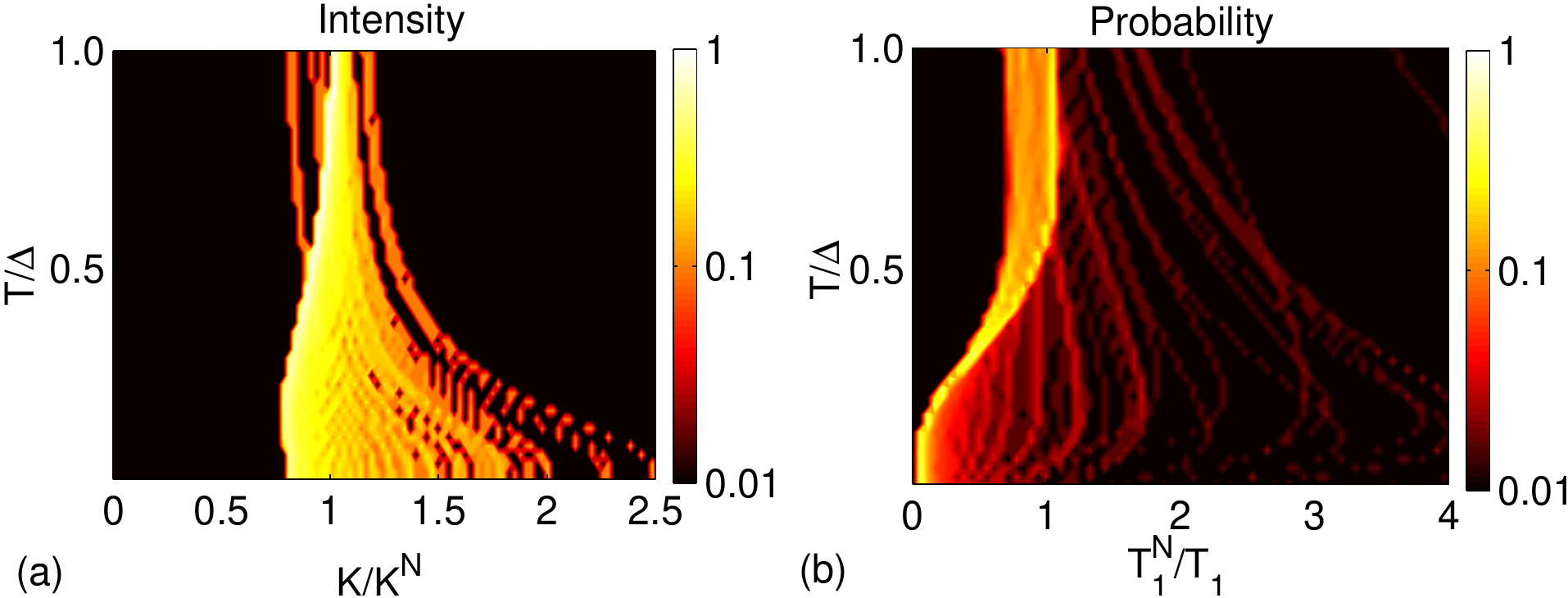}
\caption{(Color online) Distribution of (a) Knight shift and (b) spin-lattice relaxation rate for a TI film of thickness $L=240$ nm with odd-parity interorbital $\Delta$ pairing as function of temperature. The bright features are determined by the bulk, while the broad subfeatures are the surface-state response. All quantities are normalized to the normal conducting bulk at $T=0$.}
\label{fig:KT4}
\end{figure*}
%
\subsubsection{Local longitudinal spin susceptibility}

In Fig.~\ref{fig:TSC_4}(b) we plot the real part of the integrated local spin susceptibility which is proportional to the local Knight shift 
%
\begin{equation}
K(z)\sim \mathrm{Re}\int dz' \chi_{zz}(z,z';\textbf{q}_{||}=0;\omega=0).
\end{equation}
%
Here, the distinction between the bulk and the surface is even clearer than for the LDOS. Because of the unequal-spin pairing, we find a significantly reduced contribution in the bulk for $T<\Delta$. However, $K(T\rightarrow 0)$ does not vanish because of the strong spin-orbit coupling.\cite{A1959} In contrast, we find a large shift for $T\ll \Delta$ near the surface, which is even larger than the bulk shift in the normal state because of the large midgap LDOS Eq.~\eqref{eqn:Nloc}. The temperature dependence of the Knight shift is shown in Fig.~\ref{fig:KT4}(a) where the light feature is determined by the bulk and the weaker lines at larger $K$ by the surface states. The Knight shift from the surface states is spread over a wide range due to the exponential decrease in the LDOS and shows peak-like subfeatures determined by the LDOS oscillations. The lines with largest shift originate from the surface where Dirac mode and MZMs sum up, which gives rise to a very large LDOS and therefore a large spin response.

\subsubsection{Local transverse spin susceptibility}

The imaginary part of the transverse spin susceptibility Fig.~\ref{fig:TSC_4}(c) is proportional to the NMR rate \cite{HS1959}
%
\begin{equation}
\frac{1}{T_1(z)T} \sim \mathrm{Im} \sum_{\textbf{q}} \lim_{\omega \rightarrow 0} \frac{\chi_{-+}(z,z;\textbf{q}_{||};\omega)}{\omega}.
\end{equation} 
%
Similarly to the Knight shift, we clearly distinguish between the bulk and the surface. The bulk states give rise to an activation law for $T \ll \Delta$ and the Hebel-Slichter coherence peak for $T\rightarrow \Delta$. In contrast, we find a finite $T=0$ value for $1/(T_1T)$ near the surface where the contributions from the Dirac modes and the Majorana SABSs can be distinguished by their temperature behavior due to the almost constant LDOS of the Dirac modes as function of energy. This temperature dependence is shown in Fig.~\ref{fig:KT4}(b) where the surface contribution is again spread with peak-like subfeatures. However, the rate directly from the surface ($z \lesssim \xi_0$) is much larger than the rate from the MZM only ($z \gg \xi_0$), which allows us to clearly distinguish the MZM from the Dirac modes. 

In Fig.~\ref{fig:TSC_4}(d) we show the imaginary part of the dynamical transverse spin susceptibility as function of excitation energy. The spin excitation spectrum shows a very different behavior at the surface and in the bulk. For $\omega > 2\Delta E_0(k_F)/\mu$, there is a continuum of bulk spin excitations, which is sharply bounded from below. In contrast, we find edge-edge spin excitations for $\omega < 2\Delta E_0(k_F)/\mu$ at the surface and bulk-edge spin excitations for $\omega > \Delta E_0(k_F)/\mu$. However, the intensity of the edge-edge spin excitations is significantly reduced as compared to the bulk-edge spin excitations.

\subsection{Local density of states and spin response for competing pairing symmetries}

\subsubsection{Even-parity intraorbital pairing $\Delta_1$}

%
\begin{figure*}[htbp]
\centering
\includegraphics[width=1.3\columnwidth]{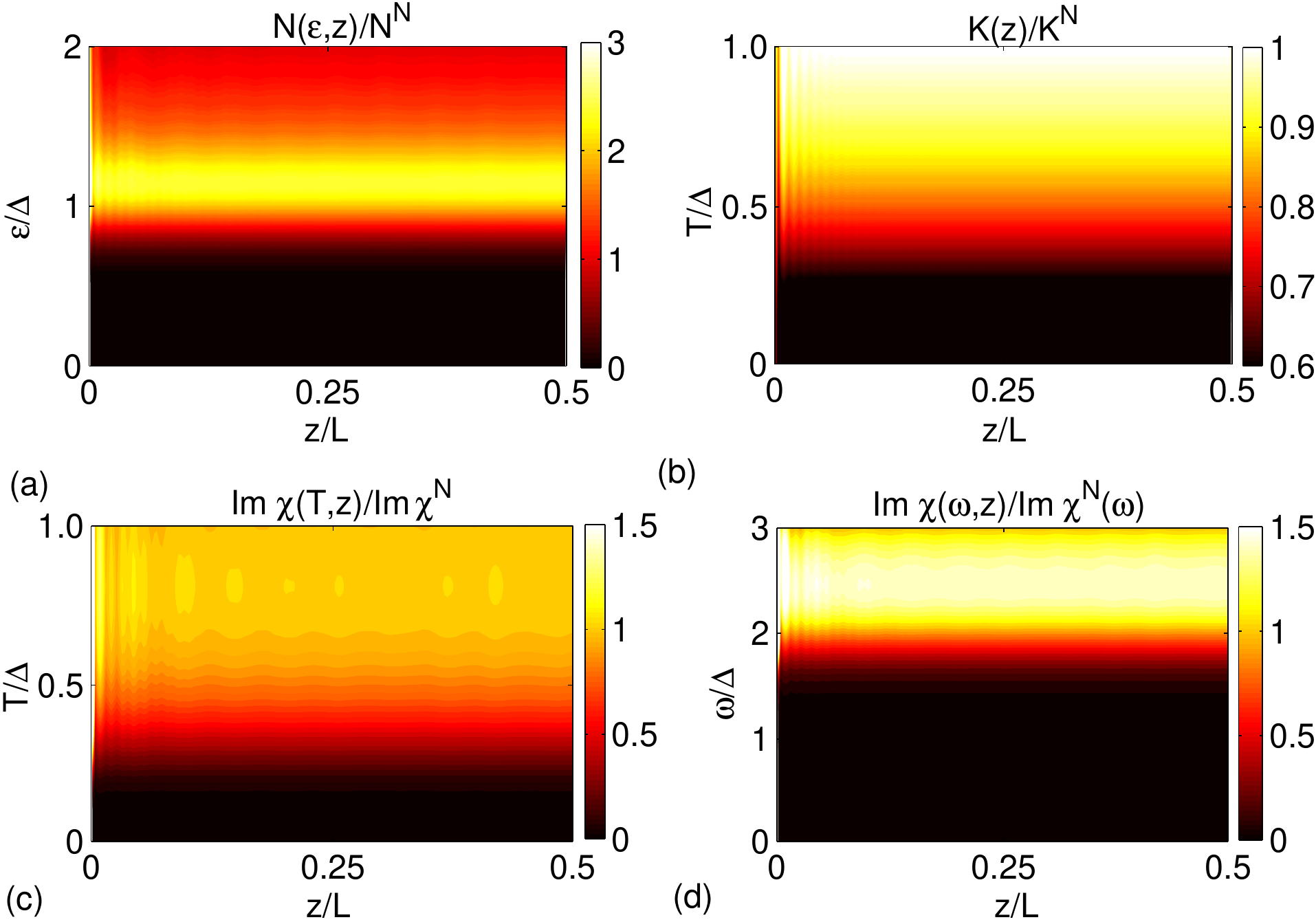}
\caption{(Color online) Spin response for a TI film of thickness $L=240$ nm and $\Delta_1$ pairing: (a) Local density of states, (b) Knight shift, and imaginary part of the transverse spin susceptibility as function of (c) temperature and (d) excitation energy. All plots are normalized to the normal conducting bulk at $T=0$. } 
\label{fig:TSC_1}
\end{figure*}
%

%
\begin{figure*}[htbp]
\centering
\includegraphics[width=1.3\columnwidth]{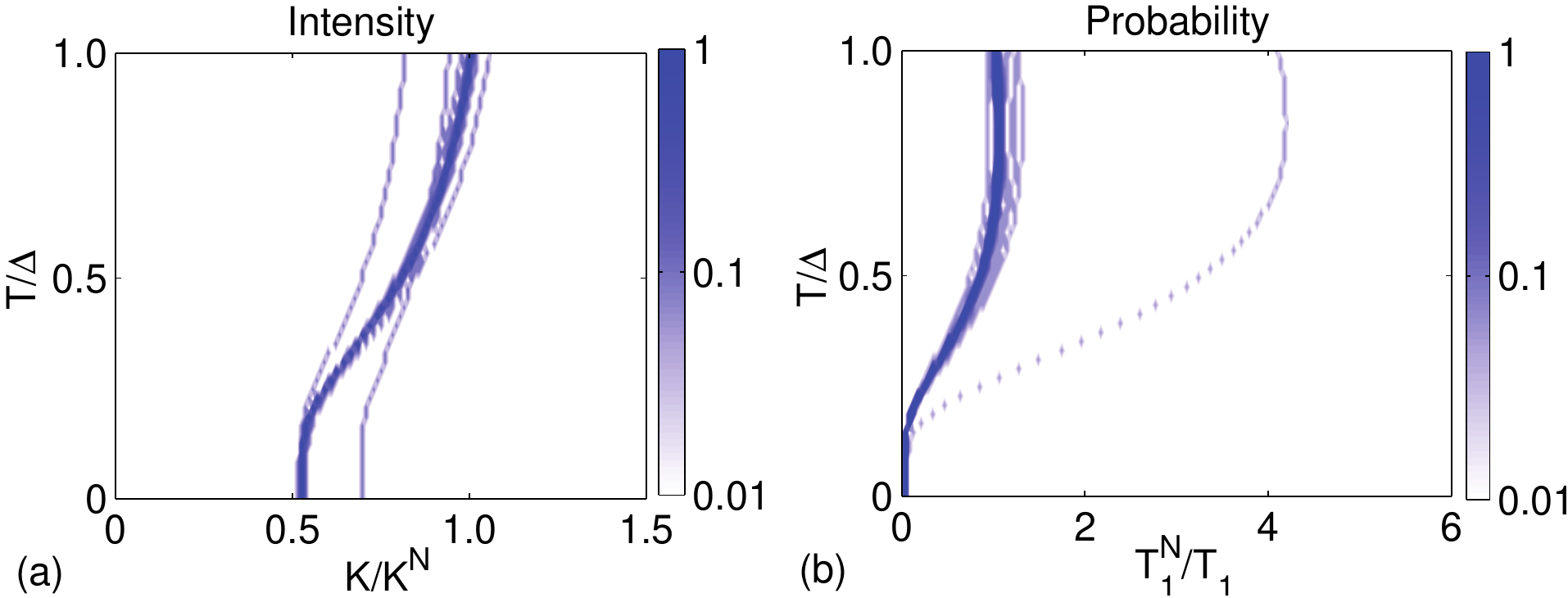}
\caption{(Color online) $\Delta_1$ pairing: Distribution of the temperature-dependent (a) Knight shift and (b) NMR rate for a TI film of thickness $L=240$. All plots are normalized to the normal conducting bulk at $T=0$. The dark features are determined by the bulk, while the subfeatures show the response from the Dirac surface modes. }
\label{fig:KT1}
\end{figure*}
%

The LDOS for $\Delta_1$ pairing is shown in Fig.~\ref{fig:TSC_1}(a) with an ordinary $s$-wave 
%
\begin{equation}
\mathcal{N}_B(\epsilon)=\frac{2\mu E_0(k_F) \epsilon}{\pi v^2v_z\sqrt{\epsilon^2-\Delta^2_1} }\Theta(\epsilon-\Delta_1)
\end{equation}
%
in the bulk. Depending on whether the Dirac modes of the TI already crossed the bulk states, we distinguish between $\mu^2>v^2|m|/B_2$ with bulk contributions only and $\mu^2<v^2|m|/B_2$ where
%
\begin{equation}
\mathcal{N}_S(\epsilon,z>0) \approx  \frac{\mu e^{-2z/\xi_0}}{\pi \xi_0 v^2}\frac{\epsilon}{\sqrt{\epsilon^2-\Delta_1^2}}\Theta(\epsilon-\Delta_1).
\label{eqn:Nloc1}
\end{equation}
%
In our numerics, we only consider the case $\mu^2>v^2|m|/B_2$ with Dirac modes. 

As shown in Fig.~\ref{fig:TSC_1}(b), the Knight shift is significantly reduced in the bulk for $T<\Delta_1$ due to the SC gap and has a finite $T=0$ value determined by the strong spin-orbit coupling. The temperature dependence of the Knight shift distribution is plotted in Fig.~\ref{fig:KT1}(a) where the dark feature is the signal from the bulk with a characteristic decrease for $T\rightarrow 0$. The line with larger shift originates from the Dirac modes and shows qualitatively the same temperature dependence as the bulk shift as expected from the gapped LDOS. 

Similarly to the Knight shift, we find conventional $s$-wave behavior for $1/(T_1T)$ in the bulk and at the surface with an activation law for $T \ll \Delta_1$ and the Hebel-Slichter coherence peak for $T\rightarrow \Delta_1$ as shown in Figs.~\ref{fig:TSC_1}(c) and \ref{fig:KT1}(b). Again, the surface shows qualitatively the same behavior as the bulk but with a much larger rate due to the large surface LDOS. The transition between the characteristic surface and bulk behaviors occurs in a depth $\xi_0$. 

In Fig.~\ref{fig:TSC_1}(d) we plot the imaginary part of the transverse spin susceptibility as function of excitation energy. For $\omega > 2\Delta_1$, there is a featureless continuum of bulk spin excitations, which is sharply bounded from below at $\omega= 2\Delta_1$ because of the quasiparticle gap. Again we find a larger spin susceptibility at the surface as compared to the bulk as a consequence of the large LDOS from the Dirac modes. 

%
\begin{figure*}[htbp]
\centering
\includegraphics[width=1.3\columnwidth]{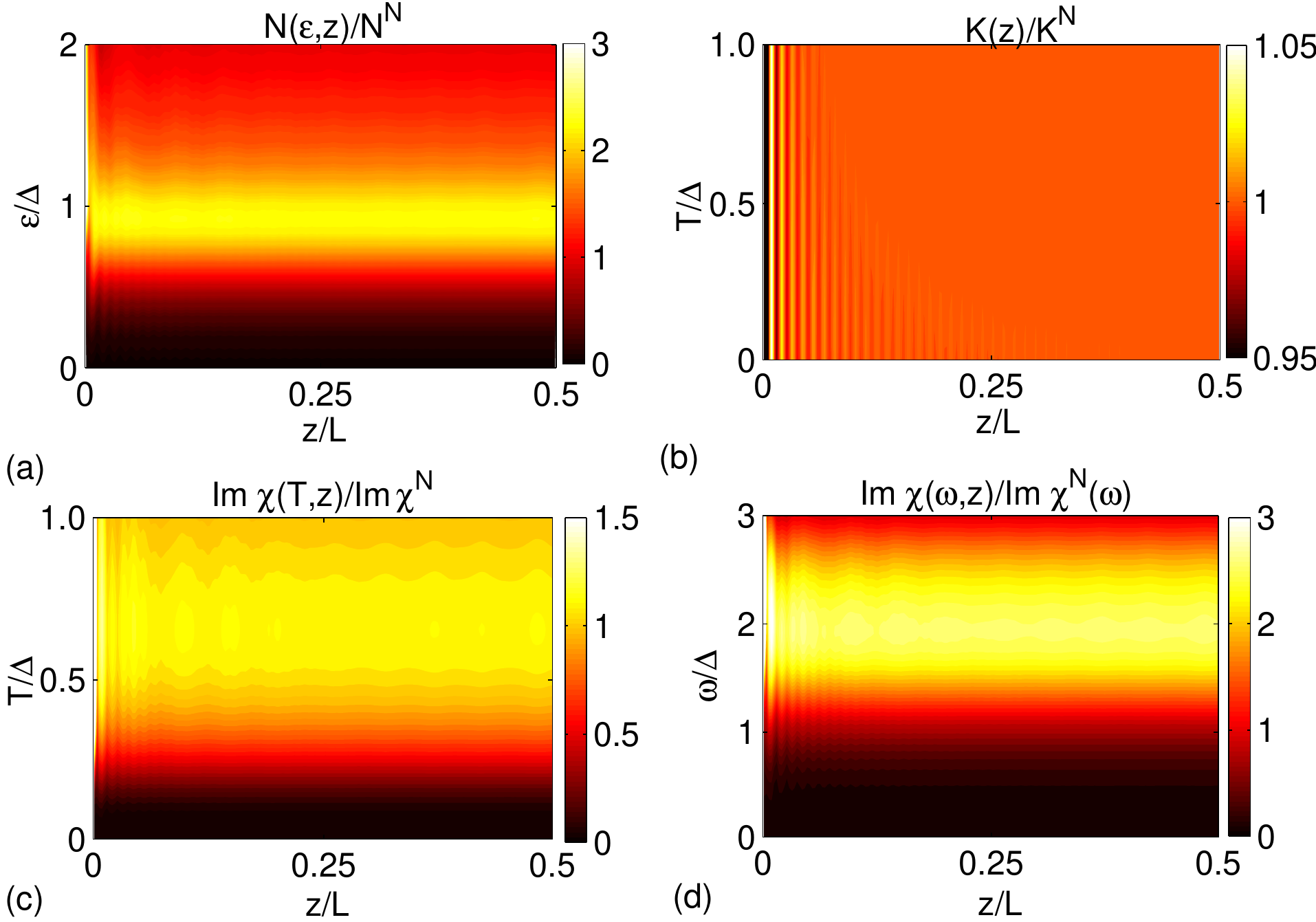}
\caption{(Color online) Spin response for a TI film of thickness $L=240$ nm and $\Delta_2$ pairing: (a) Local density of states, (b) Knight shift, and imaginary part of the transverse spin susceptibility as function of (c) temperature and (d) excitation energy. All plots are normalized to the normal conducting bulk at $T=0$. }
\label{fig:TSC_2}
\end{figure*}
%

%
\begin{figure*}[htbp]
\centering
\includegraphics[width=1.3\columnwidth]{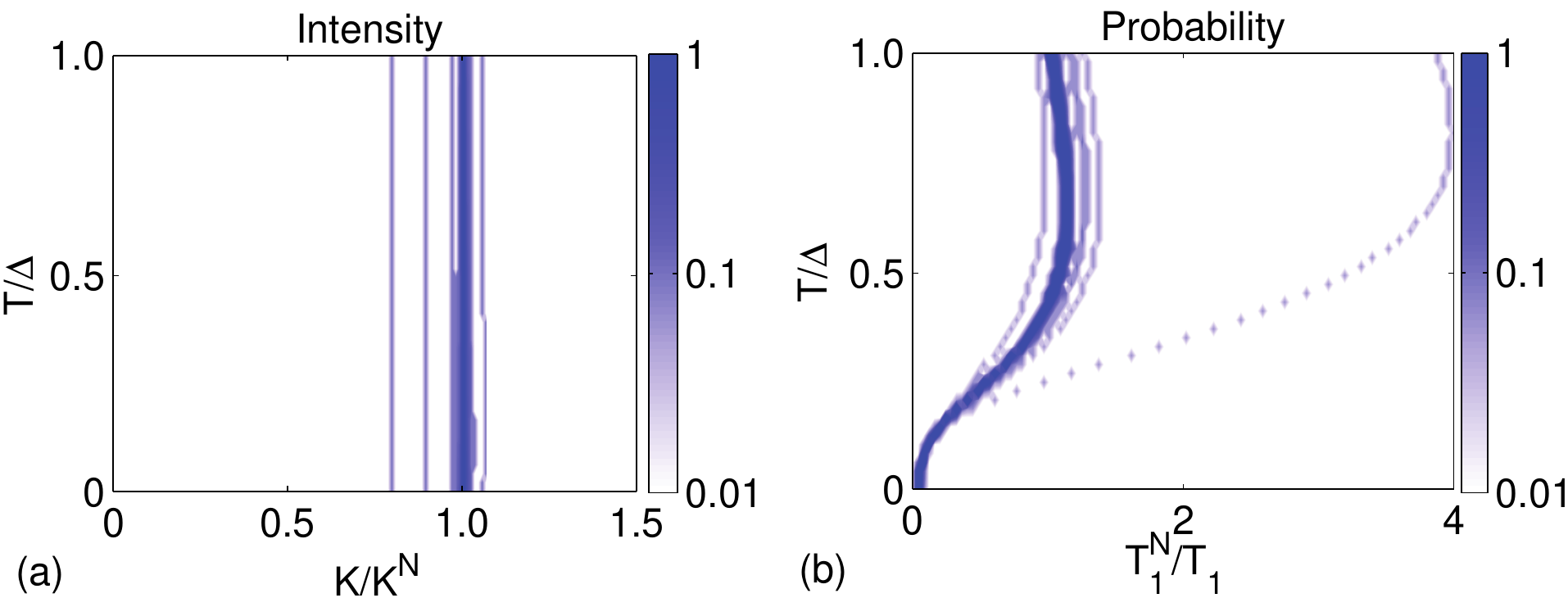}
\caption{(Color online) $\Delta_2$ pairing: Distribution of the temperature-dependent (a) Knight shift and (b) NMR rate for a TI film of thickness $L=240$ nm. All plots are normalized to the normal conducting bulk at $T=0$. The dark feature is determined by the bulk, while the subfeatures show the response from the Dirac surface modes. }
\label{fig:KT2}
\end{figure*}
%

\subsubsection{Odd-parity intraorbital pairing $\Delta_2$}

In Fig.~\ref{fig:TSC_2}(a) we plot the LDOS for $\Delta_2$ pairing, which is finite for all $\epsilon > 0$ due to the linear dispersion and shows a cusp at $\epsilon=\Delta_2$. In the bulk, we find for the LDOS the conventional result for superconductors with point nodes, 
%
\begin{equation}
\mathcal{N}_B(\epsilon)=2\frac{\epsilon \mu E_0(k_F)}{\Delta_2\pi v^2v_z}\mathrm{log}\Big|\frac{\epsilon+\Delta_2}{\epsilon-\Delta_2}\Big|.
\label{eqn:DOS2}
\end{equation}
%
As before, we distinguish between $\mu^2>v^2|m|/B_2$ with bulk contributions only and $\mu^2<v^2|m|/B_2$ with 
%
\begin{equation}
\mathcal{N}_S(\epsilon,z>0) \approx \frac{\mu e^{-2z/\xi_0}}{\pi \xi_0 v^2}\frac{\epsilon}{\sqrt{\epsilon^2-\Delta_2^2}}.
\label{eqn:Nloc2}
\end{equation}
%
The different energy dependence of the bulk and the surface dispersion yields strong evidence for different characteristic temperature behavior in the bulk and at the surface. 

As shown in Figs.~\ref{fig:TSC_2}(b) and~\ref{fig:KT2}(a), the Knight shift is essentially independent of temperature even for $T<\Delta_2$. Here, the weak subfeatures at smaller shift originate from the Dirac modes and show an $s$-wave behavior with reduced Knight shift below $T_c$ in contrast to the constant bulk value.

In Fig.~\ref{fig:TSC_2}(c) we plot the NMR rate, which is much larger at the surface than in the bulk. The temperature dependence of the rate is shown in Fig.~\ref{fig:KT2}(b) where we can clearly distinguish between bulk and surface contributions. In stark contrast to the constant Knight shift, we find for $T\rightarrow 0$ a power law $T^5$ in the bulk, which is characteristic for point nodes and an activation law $\mathrm{exp}(-\Delta_2/T)$ from the surface contribution. For $T \rightarrow \Delta_2$ we obtain a Hebel-Slichter peak at the surface whereas the bulk yields just a small coherence peak due to the broadened LDOS Eq.~\eqref{eqn:DOS2}. 

The imaginary part of the dynamical transverse spin susceptibility is shown in Fig.~\ref{fig:TSC_2}(d) as function of the excitation energy. For all $\epsilon$ there is a continuum of single-particle spin excitations with an amplitude that behaves as function of energy like $\mathcal{N}_B^2(\omega/2,z)$ in the bulk and like $[\mathcal{N}_B(\omega/2,z)+\mathcal{N}_S(\omega/2,z)]^2$ at the surface. For $\omega \approx 2\Delta_2$, there is a peak in the spin spectrum resulting from the cusp in the LDOS.

\subsubsection{Even-parity interorbital pairing $\Delta_3$}

%
\begin{figure*}[htbp]
\centering
\includegraphics[width=1.3\columnwidth]{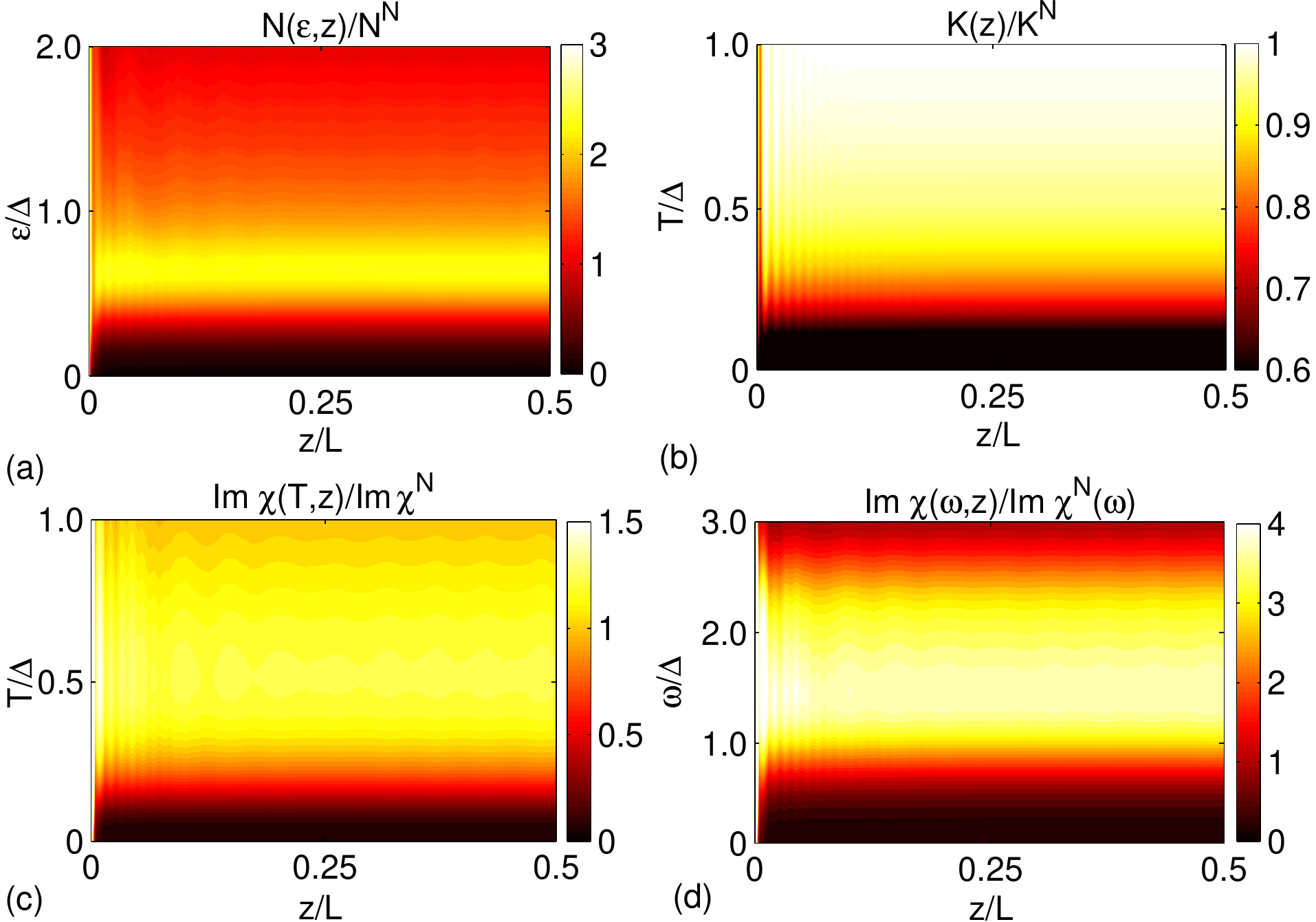}
\caption{(Color online) Spin response for a TI film of thickness $L=240$ nm and $\Delta_3$ pairing: (a) Local density of states, (b) Knight shift, and imaginary part of the transverse spin susceptibility as function of (c) temperature and (d) excitation energy. All plots are normalized to the normal conducting bulk at $T=0$. }
\label{fig:TSC_3}
\end{figure*}
%

%
\begin{figure*}[htbp]
\centering
\includegraphics[width=1.3\columnwidth]{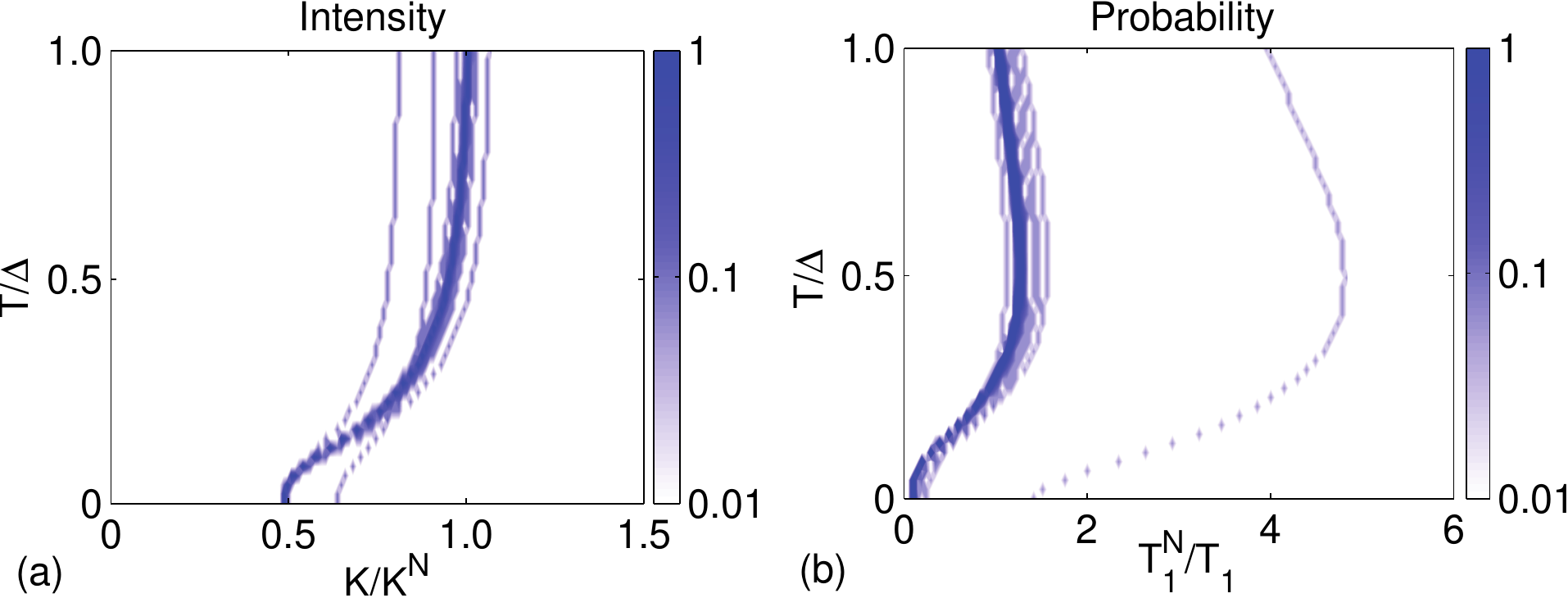}
\caption{(Color online) $\Delta_3$ pairing: Distribution of the temperature-dependent (a) Knight shift and (b) NMR rate for a TI film of thickness $L=240$ nm. All plots are normalized to the normal conducting bulk at $T=0$. The dark features are determined by the bulk, while the subfeatures show the response from the Dirac surface modes. }
\label{fig:KT3}
\end{figure*}
%

As shown in Fig.~\ref{fig:TSC_3}(a), the LDOS in the bulk 
%
\begin{equation}
\mathcal{N}_B(\epsilon)=\frac{2\mu E_0(k_F) \epsilon}{\pi v^2v_z\sqrt{\epsilon^2- \frac{\Delta_3^2m^2_0(k_F)}{\mu^2}}}\Theta \left(\epsilon-\frac{\Delta_3 |m _0(k_F)|}{\mu} \right)
\end{equation}
%
with a quasiparticle gap $2\Delta_3 |m_0(k_F)|/\mu$. From this expression, we see that the gap closes when $m_0(k)$ vanishes and decreases with increasing $\mu$. As before, we distinguish between $\mu^2>v^2|m|/B_2$ where only the bulk contributes and $\mu^2<v^2|m|/B_2$ with a gapless surface LDOS
%
\begin{equation}
\mathcal{N}_S(\epsilon,z>0) \approx  \frac{\mu e^{-2z/\xi_0}}{\pi \xi_0 v^2}.
\label{eqn:Nloc3}
\end{equation}
%

In Fig.~\ref{fig:TSC_3}(b) we plot the Knight shift $K(z)$ with qualitatively very different behavior in the bulk and at the surface. In the bulk we find a reduced shift for $T<\Delta_3$ with a finite $K(T\rightarrow 0)$ due to the strong spin-orbit coupling. In contrast, at the surface we find a larger shift for $T\ll \Delta_3$ because of the strongly spin-orbit coupled massless surface states. The temperature dependence of the Knight shift is shown in Fig.~\ref{fig:KT3}(a) where the strong feature is determined by the bulk and the second line by the Dirac modes. 

%
\begin{figure*}[htbp]
\centering
\includegraphics[width=1.3\columnwidth]{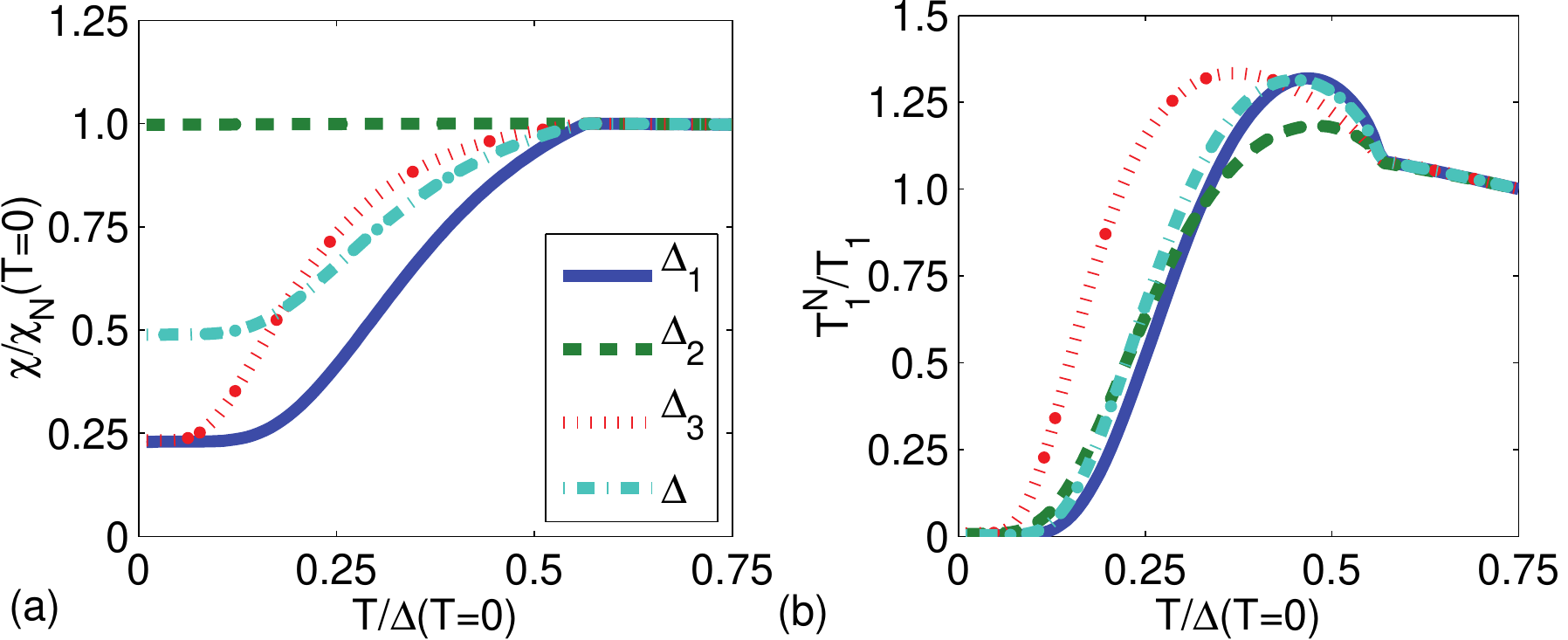}
\caption{(Color online) Temperature dependence of (a) the Knight shift and (b) the spin-lattice relaxation rate in the bulk for the different pairing symmetries with $\Delta_i=1$ meV. The lines are defined in (a). }
\label{fig:K_T}
\end{figure*}
%
In Fig.~\ref{fig:TSC_3}(c) we find similarly to the Knight shift a qualitatively different behavior for the NMR rate in the bulk and at the surface. The gapped bulk gives rise to an activation law for $T \ll \Delta_3$ and a Hebel-Slichter coherence peak for $T\rightarrow \Delta_3$. In contrast, we find a finite $T=0$-value for $1/(T_1T)$ at the surface which is characteristic for metallic states. The temperature dependence is also plotted in Fig.~\ref{fig:KT3}(b) where the rate from the Dirac modes is much larger than the rate from the bulk because of the large gapless LDOS.

In Fig.~\ref{fig:TSC_3}(d) we show the imaginary part of the dynamical transverse spin susceptibility as function of excitation energy. The excitation spectrum shows a very different behavior at the surface and in the bulk. For $\omega > 2\Delta_3|m_0(k_F)|/\mu$, there is a continuum of excitations in the bulk which is sharply bounded at $\omega= 2\Delta_3|m_0(k_F)|/\mu$. In addition, we find low-energy spin excitations at the surface and bulk-edge excitations for $\omega > \Delta_3|m_0(k_F)|/\mu $ within a length $\xi_0$ into the sample.


\section{Experimental detection scheme}
\label{sec:experimental_detection_schemes}

Finally, we discuss experimental measurement schemes of the spin susceptibility. In Refs.~\onlinecite{GH2012,TL2012,YL2012} it has been shown that $^{77}$Se and $^{209}$Bi NMR are suitable methods to investigate the bulk of Bi$_2$Se$_3$ where the dipole hyperfine coupling dominates due to the $p$-orbital character of the Fermi surface. Similarly, $^{125}$Te NMR has been successfully applied to study the metallic surface states of bismuth telluride nanoparticles.\cite{KC2013} We conclude that for low temperatures, NMR can also study the SC phase to determine the pairing symmetry. As shown in Figs.~\ref{fig:K_T}(a) and (b), we find significant differences in the bulk spin response for the competing pairing symmetries.\cite{foot_temp}

For the Knight shift, Fig.~\ref{fig:K_T}(a), we find three different behaviors below $T_c$. All pairing terms yield a finite value for the zero-temperature spin susceptibility because of the strong spin-orbit coupling. However, for $\Delta$, $\Delta_1$, and $\Delta_3$ we find a decrease of the Knight shift below $T_c$ with a similar functional behavior for $\Delta_1$ and $\Delta$, which differ by $K_N/4$ for $T=0$. For $\Delta_3$ we observe a qualitatively similar temperature dependence but the functional form is very different as for $\Delta$ and $\Delta_1$ because of the strongly momentum dependent quasiparticle gap. In stark contrast to these reduced Knight shifts, we find no change of the Knight shift for $\Delta_2$ below $T_c$. This also shows that a constant Knight shift for $T<T_c$ is not a clear signature for triplet pairing symmetry and can also appear for singlet pairing symmetries. 

In the NMR rate, Fig.~\ref{fig:K_T}(b), the differences between the pairing symmetries are not as strong as for the Knight shift. For $\Delta$ and $\Delta_1$, we find conventional $s$-wave type behavior with an activation law for $T\rightarrow 0$ and a sharp Hebel-Slichter coherence peak for $T\rightarrow T_c$. However, we can again clearly distinguish $\Delta_2$ and $\Delta_3$ from $\Delta$ and $\Delta_1$ by their functional dependence. As shown above, we find a power-law $T^5$ behavior for $\Delta_2$ due to the point nodes and for $\Delta_3$, the curve is compressed towards $T=0$ due to the momentum-dependent gap and shows no sharp coherence peak. 

Hence, bulk NMR can clearly distinguish $\Delta_2$ and $\Delta_3$ from $\Delta$, while $\Delta$ and $\Delta_1$ are qualitatively similar. Here, the main difference is that the response for $\Delta_1$ is isotropic, whereas the $\Delta$ case yields an anisotropic Knight shift.\cite{HY2012} As shown in Fig.~\ref{fig:KT4}, a more direct way to distinguish $\Delta$ and $\Delta_1$ are the SABSs for $\Delta$, which can be observed in powder samples and thin films as additional signals. The integrated intensity of the surface signal is reduced by a factor of $\xi_1/L$ compared to the bulk signal where $L$ is the thickness of the flakes. However, the surface signal shows a characteristic temperature behavior with decreasing spin susceptibility for $T\rightarrow \Delta$.

As shown in Fig.~\ref{fig:TSC_4}(c), the SABSs yield a spatially dependent relaxation rate up to $v_z/\Delta\sim 200$ nm into the sample, which is a characteristic signature for the topologically non-trivial nature of the SC state. To investigate this local magnetism, depth-resolved techniques such as $\mu$SR\cite{SB2000} and $\beta$-NMR\cite{MF2004,VF2012} are suitable. In particular, they could differentiate the regimes with one, two, or three zero-energy SABSs, which occur depending on the relative strength of the linear and quadratic terms in the Hamiltonian as discussed above. For Bi$_2$Se$_3$, it has been shown that the implanted muons most probably stop in the van der Waals gap between quintuple layers.\cite{SP2012} Despite our concentration on NMR and $\mu$SR, our results are more general and can also be applied to electron spin resonance and surface sensitive spin-flip Raman scattering.\cite{PA2007,HP2011}

Doped TIs usually have SC shielding fractions of 50\% while the rest is normal conducting.\cite{SK2011} Here, bulk NMR could detect the mechanisms, which determine this fraction. Depending on $\mu$ in the normal part of the sample, there might be additional signals, which can be clearly distinguished from the SC ones by the very different temperature behavior. This key information about the distribution of dopants is an important step towards the understanding of unconventional superconductivity in TIs. In this paper, we did not take vortices into account, which can be distinguished from the SABS by their temperature behavior and the different length scales.


\section{Conclusion}
\label{sec:conclusion}
In conclusion, we have proposed a NMR experiment to determine the pairing symmetry of doped three-dimensional TIs and to observe the MZMs in the odd-parity interorbital unequal-spin pairing channel. Quadratic momentum terms significantly determine the character of the different species of unconventional SABSs. We have shown that the surface states of these systems have two characteristic length scales originating from the band inversion and the odd-parity pairing, and typically differ by two orders of magnitude. They contribute to a local spin susceptibility, which can be clearly distinguished from the bulk by their characteristic temperature behavior and the large local density of states. Furthermore, we emphasize the usefulness of depth-controlled local probes, which directly show the existence of unconventional SABSs. 

We acknowledge N. Georgieva, J. Haase, and D. Rybicki for useful discussions, and the Federal Ministry of Education and Research (BMBF) for financial support.


\begin{thebibliography}{99}
\bibitem{foot_temp} In this section we assume that the order parameter $\Delta_i$ is temperature dependent with $\Delta_i(T)=\Delta_i\sqrt{1-T/1.76\Delta_i}$.
\end{thebibliography}
\end{document}